\documentclass[pre,twocolumn,english,aps,prb,a4paper,floatfix]{revtex4}
\usepackage[T1]{fontenc}
\usepackage{color}
\usepackage{amsmath}
\usepackage{amssymb}
\usepackage[latin1]{inputenc}
\usepackage{graphicx}
\usepackage{bm}
\usepackage{epsfig}
\usepackage{float}
\DeclareMathOperator{\Tr}{Tr}

\begin{document}

\title{Phase behaviour and correlations of parallel hard squares: From highly confined to bulk systems}

\author{Miguel Gonz\'alez-Pinto}
\email{miguel.gonzalezp@uam.es}
\address{Departamento de F\'{\i}sica Te\'orica de la Materia Condensada, Facultad de Ciencias, 
Universidad Aut\'onoma de Madrid, E-28049 Madrid, Spain}

\author{Yuri Mart\'{\i}nez-Rat\'on}
\email{yuri@math.uc3m.es}
\address{Grupo Interdisciplinar de Sistemas Complejos (GISC), Departamento de Matem\'aticas, 
Escuela Polit\'ecnica Superior, Universidad Carlos III de Madrid, Avenida de la Universidad 
30, E-28911, Legan\'es, Madrid, Spain}

\author{Szabolcs Varga} 
\email{vargasz@almos.uni-pannon.hu}
\author{Peter Gurin} 
\email{gurin@almos.uni-pannon.hu}
\address{Institute of Physics and Mechatronics, University of Pannonia, 
PO BOX 158, Veszpr\'em, H-8201 Hungary}

\author{Enrique Velasco}
\email{enrique.velasco@uam.es}
\address{Departamento de F\'{\i}sica Te\'orica de la Materia Condensada, Instituto 
de Ciencia de Materiales Nicol\'as Cabrera and IFIMAC, Universidad Aut\'onoma de Madrid, E-28049 Madrid, Spain}

\begin{abstract}
We study a fluid of two-dimensional parallel hard squares in bulk and under confinement in channels, with the aim of evaluating the 
performance of Fundamental-Measure Theory (FMT). To this purpose, we first analyse the phase behaviour 
of the bulk system using FMT and Percus-Yevick theory, and compare the results with MD and MC 
simulations. In a second step, we study the confined system and check the results against those obtained 
from Transfer Matrix Method and from our own Monte Carlo simulations. 
Squares are confined to channels with parallel walls at angles of 0$^{\circ}$ or 45$^{\circ}$ relative to the diagonals of the 
parallel hard squares, respectively,
which allows for an assessment of the effect of the external-potential symmetry 
on the fluid structural properties.  
In general FMT overestimates bulk correlations, predicting the existence of a columnar phase (absent in simulations) prior to 
crystallisation. The equation of state predicted by FMT compares well with simulations, although the PY approach with the virial route 
is better in some range of packing fractions. 
The FMT is highly accurate for the structure and 
correlations of the confined fluid due to the dimensional crossover property fulfilled by the theory.
Both density profiles and equations of state of the confined system are accurately predicted by the theory. 
The highly non-uniform pair correlations inside the channel are also very well described by FMT.  
\end{abstract}

\date{\today}

\maketitle

\section{Introduction}
\label{intro}

Density functional theory (DFT) has proved to be a very successful tool to predict the phase behaviour
of bulk and confined classical fluids \cite{Evans,Hansen}. Since the local-density approximation is not appropriate for
classical fluids, early versions of DFT for the hard-sphere (HS) system included correlations through averaged local densities, 
the effective density approximation \cite{Lutsko} or the weighted density approximation \cite{Curtin,Tarazona1} being two
widely used versions. Since then, DFT for HS has evolved to converge to a more sophisticated class of approximations: 
the so-called fundamental measure 
density functional theory (FMT). This theory was proposed by Rosenfeld in the 80's \cite{Yasha1,Yasha2}, went through a period of
refinement (in an effort to adequately describe HS crystallization \cite{Schmidt1}), and current versions adequately describe
crystal anisotropies at high densities \cite{Tarazona2} or the HS equation of state (EOS) for fluid phases \cite{Roth1,Roth1b}.
Competent reviews of FMT for mixtures of HS and other hard particle systems can be found in \cite{Roth2}
and \cite{Tarazona3} respectively.

The first FMT functional for anisotropic particles was developed for mixtures of parallel hard squares (PHS) in 2D, mixtures of parallel
hard cubes (PHC) in 3D, and also for a ternary mixture of hard rectangles (2D) or parallelepipeds (3D)
with restricted orientations (Zwanzig approximation)
\cite{Cuesta0,Cuesta1,Cuesta2}. These density functionals were used to calculate the phase diagrams of the one-component fluid, binary mixtures
of hard cubes \cite{Yuri1}, and also prolate and oblate Zwanzig particles \cite{Yuri2}. Recently they were
also applied to the study of the phase behaviour of hard biaxial board-like particles \cite{Yuri3},
polydisperse mixtures of highly oriented hard platelets \cite{Velasco1}, and Zwanzig particles confined in a square cavity \cite{Miguel} or in geometrically structured three-dimensional surfaces \cite{Harnau}.

FMT density functionals were also obtained for binary or ternary mixtures
of freely rotating needles, platelets of vanishing thickness, and HS \cite{Schmidt2,Schmidt3,Schmidt4}. These
functionals were applied to study the demixing behavior \cite{Schmidt2} and more recently the stacking phase diagrams
of binary mixtures of anisotropic particles \cite{Heras}. A FMT functional for mixtures of parallel hard cylinders of finite thickness has been obtained within the 
dimensional crossover property \cite{Yuri_cylinders}. More recently, numerically-tractable versions of FMT functionals were obtained 
for freely rotating anisotropic particles which exploited the approximate decomposition of the Mayer function
as convolutions of one-particle weights \cite{Mecke1,Mecke2,Mecke3}, an idea originally 
proposed by Rosenfeld \cite{Rosenfeld_Mayer}. These versions were successful in the analysis of structural properties of
platonic solids in contact with hard walls \cite{Marechal1}, and in the study of the bulk phase behaviour of hard spherocylinders including also the smectic phase
\cite{Mecke2}. For a recent review on DFT applied to the study of hard body models see Ref. \cite{Mederos}.

It is usually accepted that a density functional fulfilling the dimensional crossover property should provide accurate predictions for the
structure of highly confined fluids. The dimensional crossover property means that a functional for $D$-dimensional particles reduces to that for
$D-1$-dimensional particles if density profiles are constrained from the higher to the lower dimension, provided both functionals were obtained
separately from the same formalism. With this property alone FMT functionals for HS and hard disks can be obtained \cite{Tarazona_0D}. 
The FMT functionals were proved to be very accurate in the description of HS in high confinement 
\cite{White1,White2,Wu,Mansoori,Mariani}, but there is not enough evidence of that for  
other anisotropic particles. Also, the FMT accurately predicts the properties of HS crystals 
\cite{Lutsko2} and of the fluid-crystal interface \cite{Hartel,Oettel}.

\begin{figure}
\epsfig{file=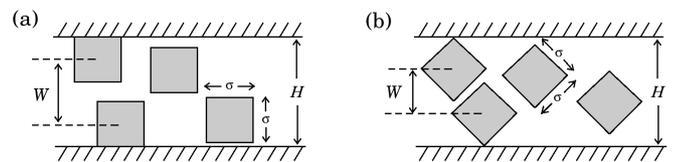,width=3.5in}
\caption{Schematic of the two channels studied. (a) A channel with walls parallel to one of the sides of the squares. (b) A channel with walls at an angle
of 45$^{\circ}$ with respect to the particle sides. $\sigma$ is the side-length of the squares. $H$ is the channel thickness. $W$ is the length available
to the centres of mass of the squares in the transverse direction. The relation between $H$ and $W$ is $H=W+\sigma$ in (a), and $H=W+\sigma\sqrt{2}$ in (b).}
\label{fig1m}
\end{figure}

In the present article we study
the performance of FMT in the description of two-dimensional fluids of confined PHS. Even though hard discs (HD) may be considered to be geometrically
simpler than PHS at first sight, in fact the dimensional-crossover-compliant FMT functional of HD contains a complicated two-body weighted density,
in contrast with that of PHS, which features only one-body weighted densities.
We numerically implement the FMT functional for PHS to study the
thermodynamics (EOS), structure (density profiles) and correlation (pair correlation functions) of the confined fluid and
check these results against transfer matrix method (TMM) and our own Monte Carlo (MC) simulations \cite{MC}. Particles are confined in a narrow channel with parallel hard walls,
such that only two particles can fit in the transverse direction of the channel, Fig. \ref{fig1m}.
Two different channels, corresponding to two different symmetries of the external potential representing the walls, will be studied: (i) a channel with
walls parallel to one of the sides of the PHS, Fig. \ref{fig1m}(a), and (ii) a channel with walls at an angle of 45$^{\circ}$ with respect to the 
particle sides, Fig. \ref{fig1m}(b). The results presented here confirm the expectation that the
FMT functional accurately describes the structure of highly confined fluids.

However one could expect that, as the channel thickness becomes larger and the bulk limit is approached, the results will become progressively
worse. For the purpose of evaluating the predictive power of the present functional in the description of the bulk system, we performed a minimisation
using a Gaussian parametrization (note that a free-minimization was recently performed for the same functional in Ref. \cite{Roij1}, which
concluded that the Gaussian parametrization accurately describes the EOS and the phase transitions), and check the resulting EOS
against MD simulations \cite{Hoover_MD}. Alternatively, the EOS from the Percus-Yevick (PY) approximation, both from virial and compressibility routes,
were also obtained and compared with simulations. Finally, pair correlations functions were calculated from (i) the same PY
approximation, (ii) from the Ornstein-Zernike relation together with the direct correlation function obtained from the FMT functional, and 
(iii) from the test-particle route (which involves functional minimisation with a particle fixed at the origin).
Apart from predicting a spurious columnar (C) phase (already reported in \cite{Roij1,Miguel}), and overemphasising pair correlations, the agreement
between FMT and simulations is acceptable, especially regarding the EOS at high densities and the prediction of a relatively high percentage of 
vacancies in the crystal (K) phase, an issue recently confirmed by simulations \cite{Dijkstra1,Marechal2}.

Finally, we would like to motivate the use of simple hard models in low and restricted geometries. In principle, 
the hard-square and hard-cube systems can be simply regarded as purely academic models where the tools of statistical mechanical can 
be applied with great simplicity \cite{Hoover1,Hoover2}. However, advances in the design and synthesis of colloidal particles of different
shapes have allowed the realisation of experiments on colloidal hard cubes (with rounded edges \cite{holandeses1,holandeses2}). Colloidal particles
can be nano-confined with the help of external potentials to form a single monolayer. This was done recently with
hard discs standing on edge \cite{Chaikin}: the system so obtained behaved like freely-rotating hard rectangles which formed a nematic
phase with tetratic symmetry \cite{Chaikin}, a phase also found in simulations \cite{Frenkel,Torquato} .
Recent experiments on monolayers of hard cubes with rounded edges observed
hexagonal rotator and rhombic crystal phases \cite{chinos}, whose stability was also confirmed by simulations on two-dimensional
hard squares with rounded corners \cite{Escobedo}.  When the degree of roundness is small, the monolayer can be viewed
as an effective two-dimensional hard-square fluid, which at high packing fractions behaves like a PHS system.

\section{Fundamental measure density functional for parallel hard squares}

In this section we briefly describe how to obtain the FMT functional for a mixture of PHS. For more details see Refs. \cite{Cuesta1} and 
\cite{Cuesta2}. The formalism used here can be applied in any dimension, but the main ingredients are already present in two dimensions. 
The Mayer function for two PHS of edge-lengths $\sigma_{\mu}$ and $\sigma_{\nu}$ parallel to the $x$ and $z$ axes 
can be decomposed as a sum of convolutions: 
\begin{eqnarray}
&&-f_{\mu\nu}({\bm r})=\Theta\left(\frac{\sigma_{\mu}+\sigma_{\nu}}{2}-|x|\right)
\Theta\left(\frac{\sigma_{\mu}+\sigma_{\nu}}{2}-|z|\right)\nonumber\\
&&=\left\langle\omega^{(0)}_{\mu}\ast\omega^{(2)}_{\nu}\right\rangle({\bm r})+
\left\langle\omega^{(1x)}_{\mu}\ast\omega_{\nu}^{(1z)}\right\rangle({\bm r}),
\label{mayer}
\end{eqnarray}
where $\Theta(x)$ is the Heaviside function. We use the notation $\langle g_{\mu\nu}\rangle=g_{\mu\nu}+g_{\nu\mu}$, 
i.e. the symmetrization with respect to the 
particle indexes, and the one-particle weights $w^{(\alpha)}_{\mu}({\bm r})$ are defined as 
\begin{eqnarray}
\omega^{(0)}_{\mu}({\bm r})&=&\frac{1}{4}\delta\left(\frac{\sigma_{\mu}}{2}-|x|\right)\delta\left(\frac{\sigma_{\mu}}{2}-|z|\right), 
\label{w0}\\
\omega^{(1x)}_{\mu}({\bm r})&=&\frac{1}{2}\Theta\left(\frac{\sigma_{\mu}}{2}-|x|\right)\delta\left(\frac{\sigma_{\mu}}{2}-|z|\right),\\
\omega^{(1z)}_{\mu}({\bm r})&=&\frac{1}{2}\delta\left(\frac{\sigma_{\mu}}{2}-|x|\right)\Theta\left(\frac{\sigma_{\mu}}{2}-|z|\right),\\
\omega^{(2)}_{\mu}({\bm r})&=&\Theta\left(\frac{\sigma_{\mu}}{2}-|x|\right)\Theta\left(\frac{\sigma_{\mu}}{2}-|z|\right),
\label{w2}
\end{eqnarray}
with $\delta(x)$ the Dirac-delta function.   
Using these weights, we define the weighted densities: 
\begin{eqnarray}
n_{\alpha}({\bm r})=\sum_{\mu} \left[\rho_{\mu}\ast \omega^{(\alpha)}_{\mu}\right]({\bm r}),
\label{weigthed}
\end{eqnarray}
so that the low-density expansion of the excess free-energy functional, with second functional derivatives given 
by (\ref{mayer}), is
\begin{eqnarray}
\beta {\cal F}_{\rm exc}[\{\rho_{\mu}\}]\rightarrow
\int d{\bm r}\left[n_0({\bm r})n_2({\bm r})+ n_{1x}({\bm r})n_{1z}({\bm r})\right],
\label{low}
\end{eqnarray}
($\beta^{-1}\equiv k_B T$) which in turn constitutes the second virial expansion of the FMT functional we are looking for. 

Second, the excess free-energy density of a uniform mixture of PHS according to scaled-particle 
theory (SPT) \cite{Reis} is
\begin{eqnarray}
\Phi=-\xi_0 \log(1-\xi_2)+\frac{\xi_1^2}{1-\xi_2},
\end{eqnarray}
where
\begin{eqnarray}
\xi_0=\sum_{\mu}\rho_{\mu}\cdot 1, \quad \xi_1=\sum_{\mu}\rho_{\mu}\cdot \sigma_{\mu},\quad 
\xi_2=\sum_{\mu}\rho_{\mu}\cdot \sigma_{\mu}^2,
\end{eqnarray}
are sums of the number densities multiplied by the fundamental measures of 
particles. The spatial integrals of the weight functions (\ref{w0})-(\ref{w2})
produce the same fundamental measures:
 $\int d{\bm r} \omega^{(\alpha)}_{\mu}({\bm r})=\sigma_{\mu}^{\alpha}$ ($\alpha=0,2$), and 
$\int d{\bm r} \omega^{(\beta)}_{\mu}({\bm r})=\sigma_{\mu}$ ($\beta=1x,1z$). Therefore, identifying the weighted densities 
of non-uniform mixtures with their values $\xi_i$ for uniform density
$\xi_0\to n_0({\bm r})$, $\xi_{1}^2\to n_{1x}({\bm r})n_{1z}({\bm r})$ and $\xi_2\to n_2({\bm r})$, we finally obtain 
the excess free-energy functional whose low-density expansion is given by (\ref{low}):  
\begin{eqnarray}
&&\beta{\cal F}_{\rm exc}[\{\rho_{\mu}\}]=\int d{\bm r} \Phi({\bm r})\nonumber\\
&&=\int d{\bm r}\left[-n_0({\bm r})\log(1-n_2({\bm r}))+\frac{n_{1x}({\bm r})n_{1z}({\bm r})}{1-n_2({\bm r})}\right],
\nonumber\\
\label{df}
\end{eqnarray}
An important property fulfilled by this functional is the dimensional crossover, i.e. when the particle
centres of mass are constrained to the $x$ axis, i.e. $\rho_{\mu}({\bm r})=\rho_{\mu}(x)\delta(z)$,
the exact Percus density-functional for hard segments is recovered:
\begin{eqnarray}
&&\beta {\cal F}_{\rm exc}[\{\rho_{\mu}\}]\Big|_{\rho_{\mu}({\bm r})=\rho_{\mu}(x)\delta(z)}\nonumber\\
&&=-\int dx 
\tilde{n}_0(x)\log (1-\tilde{n}_1(x)),  
\end{eqnarray}
with 
\begin{eqnarray}
&&\tilde{n}_0(x)=\frac{1}{2}\sum_{\mu} \left[\rho_{\mu}(x-\sigma_{\mu}/2)+\rho_{\mu}(x+\sigma_{\mu}/2)\right],\\
&&\tilde{n}_1(x)=\sum_{\mu} \int_{x-\sigma_{\mu}/2}^{x+\sigma_{\mu}/2}dx'\rho_{\mu}(x'),
\end{eqnarray}

In the present work we use (\ref{df}) with the weighted densities (\ref{weigthed}) for the one-component case 
(i.e. there is only one density profile $\rho({\bm r})$) as the FMT excess density functional. The ideal part is, as usual 
\begin{eqnarray}
\beta {\cal F}_{\rm id}[\rho]=\int d{\bm r} \rho({\bm r})\left[\log \rho({\bm r})-1\right].
\end{eqnarray}
The total free-energy density functional ${\cal F}[\rho]={\cal F}_{\rm id}[\rho]+{\cal F}_{\rm exc}[\rho]$ is minimized at fixed mean 
density $\rho_0$ to obtain the equilibrium density profiles for the non-uniform phases in bulk. For the confined fluids one minimises
the grand potential,
\begin{eqnarray}
\Omega[\rho]={\cal F}[\rho]+\int d{\bm r} \rho({\bm r})\left[v_{\rm ext}({\bm r})-\mu\right],
\end{eqnarray}
where $\mu$ is the bulk chemical potential, while $v_{\rm ext}({\bm r})$ is the external potential acting on the 
particle centres of mass.
Also we calculate the longitudinal pressure of the confined fluid as 
\begin{eqnarray}
&&\beta p =A^{-1}\left[\int_A d{\bm r} \rho({\bm r}) \frac{\delta \beta {\cal F}[\rho]}{\delta \rho ({\bm r})}-\beta{\cal F}[\rho]\right]
\nonumber\\&&=A^{-1}\int_A d{\bm r} \left[\frac{n_0({\bm r})}{1-n_2({\bm r})}+\frac{n_{1x}({\bm r})n_{1y}({\bm r})}
{(1-n_2({\bm r}))^2}\right],
\end{eqnarray}
where $A$ is the total area of the box. 

We will consider two different external potentials $v_{\rm ext}({\bm r})$, reflecting two different symmetries for the channel:
\begin{itemize}
\item[(i)] The hard walls of the channel are parallel to one of the sides of the squares along the $x$ axis, Fig. \ref{fig1m}(a): 
\begin{eqnarray}
\beta v_{\rm ext}({\bm r})=
\left\{
\begin{matrix}
0, & \text{if}\ \displaystyle{|z|\leq \frac{H-\sigma}{2}},\\
\infty, & \quad \text{otherwise}.
\end{matrix}
\right.
\end{eqnarray}
\item[(ii)] The walls are at an angle of 45$^{\circ}$ with respect to any of the two particle sides, Fig. \ref{fig1m}(b):
\begin{eqnarray}
\beta v_{\rm ext}({\bm r})=
\left\{
\begin{matrix}
0, & \text{if}\ \displaystyle{|z-x|\leq\frac{H}{\sqrt{2}}-\sigma},\\
\infty, & \quad \text{otherwise}.
\end{matrix}
\right.
\end{eqnarray}
\end{itemize}
$H$ is the thickness of the channel.
For the parallel case we will consider channels with $2\sigma\leq H< 3\sigma$, while for the 
oblique case $\sqrt{2}\sigma\leq H< 3\sqrt{2}\sigma$; in both cases the free area available to squares only allow two particles to
fit along the transverse direction.

\section{Transfer matrix theory for confined fluids}

As mentioned in Sec. \ref{intro} we have focused our interest on channels that can hold two particles at most in the transverse direction, 
mainly because for these small systems the TMM can produce exact results.
It is still a great challenge to extend the list of exactly solvable models in the
direction of confined fluids starting from the one-dimensional Tonks-gas \cite{Tonks}, where the
particles interacts with only few neighbours. The problem arises from the appearance of
additional positional freedoms, which gives rise to increasing number of interactions.
Therefore the successful one-dimensional methods must be extended for pair interactions
with a finite number of neighbours. Along this line the transfer matrix method has been
proved very successful to get exact results in very narrow pores \cite{Kofke}.

Even if the formalism of the transfer matrix method is transparent the
equations to obtain the eigenfunctions and eigenvalues (from which we can
obtain the thermodynamical and structural properties of the fluid) can be
obtained in analytic form only for few model systems such as the classical
spin models on one and two-dimensional lattices \cite{Joyce,Yeomans}, one-dimensional
gas of some rotors \cite{Casey,Gurin1} and the quasi-one-dimensional fluid of
hard rhombuses in very narrow pores \cite{Varga}. However, the equation for
eigenfunctions has been solved numerically for several continuum
models in the past, where only nearest neighbour interactions are present. The most
important examples are the system of hard disks confined between two parallel walls
\cite{Kofke,Varga,Godfrey1,Ashwin,Yamchi} and that of hard spheres in cylindrical
channel \cite{Kofke,Kamenetskiy,Gurin2}. The application of the transfer matrix method
has proved to be a formidable task for those systems, where the particles can
interact with $m>1$ neighbours. The first numerical transfer matrix solution for
parallel hard squares confined between soft walls (periodic boundary walls), where
both first and second neighbour interactions are present, is due to Percus and
Zhang \cite{Percus}. Recently the method has been applied successfully in the
presence of hard walls for both hard disks (HD) \cite{Godfrey2} and PHS
\cite{Gurin3}.

\begin{figure}
\epsfig{file=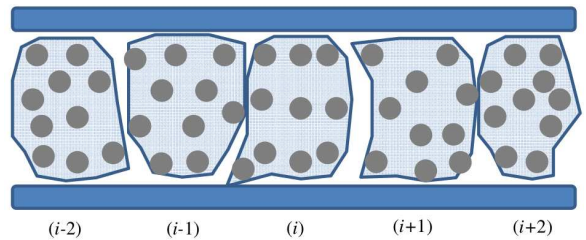,width=3.in}
\caption{The smallest separable clusters ($m$-mer) of the confined fluids.}
\label{fig0}
\end{figure}

The basic idea of TMM is to find the configurational part of the partition function as
a trace of the matrix products of low dimensional matrices. In the case of 2D
confinement, where particles are confined between two parallel walls (see Fig. \ref{fig0}), it is
useful to change from the canonical to the isobaric ensemble, where the longitudinal
pressure $p$ is chosen to be the independent variable. In this new ensemble, one can
perform changes of variables in particle positions and find the size of the minimal cluster
($m$) to simplify the isobaric partition function as much as possible. 

Without going into details, one can show that the configurational
part of the partition function can be written as
\begin{eqnarray}
&&   Z_{NpT}
  =  \int d\chi_1 \cdots \int d\chi_{N/m} \;
     K(\chi_1,\chi_2) \dots  K(\chi_{N/m},\chi_1)\nonumber\\
 && =  \Tr(K^{N/m})
     \; ,
\label{partition}
\end{eqnarray}
where the kernel function is defined as
\begin{equation}
     K(\chi_i,\chi_{i+1})
  =  \int_0^\infty e^{-\beta[U(\chi_i,\chi_{i+1},X_{i,i+1})+pHX_{i,i+1}]} dX_{i,i+1}
     \; .
 \label{eq:K}
\end{equation}
Here $\chi_i$ is a short notation for all the internal positional
variables of the $i$-th cluster, i.e. the $m$ pieces of transverse
coordinates and $m-1$ pieces of relative longitudinal coordinates of
the $m$ particles of the $i$-th cluster. Furthermore, $X_{i,i+1}$ is
the difference between the longitudinal coordinates of the centres of
the neighbouring ($i$-th and $i+1$-th) clusters,
$U(\chi_i,\chi_{i+1},X_{i,i+1})$ is the whole potential energy of
particles form the $i$-th and $i+1$-th clusters, which is infinity if
any two particles overlap and zero otherwise.  We also mention
that the matrix product is defined as $K^2(\chi_{i-1},\chi_{i+1}) =
\int d\chi_i K(\chi_{i-1},\chi_{i}) K(\chi_{i},\chi_{i+1})$.

In the derivation of Eqn. (\ref{partition}) we 
emphasize that the size of the minimal cluster is equal to the number of 
possible interactions of a chosen particle with the other ones if the 
positional order of the particles are kept fixed along the longitudinal 
direction ($x_1<x_2\cdots<x_N$). This is equivalent to the statement that 
all particles can interact with $2m$ neighbours only ($m$ neighbours in 
forward and $m$ neighbours in backward directions along the pore). In addition to this, the $\chi_1$ 
and $\chi_{N/m+1}$ coordinates are the same in (\ref{partition}). The size 
of the minimal cluster depends on the particle-particle interaction and 
the width of the channel. Fig. \ref{fig0} represents a case where a system can 
be divided into clusters of ten particles. Since the result of the trace 
operation in Eqn. (\ref{partition}) does not depends on the basis used, it is worth 
determining the eigenvalues and the eigenfunctions of the kernel function 
($K$) to obtain the partition function, as follows:
\begin{eqnarray}
Z_{NpT}=\text{Tr}\left(K^{N/m}\right)=\sum_n \lambda_n^{N/m},
\end{eqnarray}
where $\lambda_n$ and $\Psi_n$ are the $n$th eigenvalue and eigenfunction 
of the following integral equation:
\begin{eqnarray}
\int d\chi_{i+1}K(\chi_i,\chi_{i+1})\Psi_n(\chi_{i+1})=\lambda_n
\Psi_n(\chi_i).
\end{eqnarray}
As can be seen from Eq. (\ref{eq:K}), the kernel function is not
symmetric for wider channels because a cluster formed by more than one
particle is generally not symmetric under reflection, and if we
interchange the $i$-th and $i+1$-th clusters the opposite sides of the
clusters will interact with each other. Therefore we can get a second
eigenvalue equation in the following form: 
\begin{eqnarray}
\int d\chi_iK(\chi_i,\chi_{i+1})\overline{\Psi}_n(\chi_i)=
\overline{\lambda}_n\overline{\Psi}_n(\chi_{i+1}).
\label{integral}
\end{eqnarray}
Note that it can be proved that $\overline{\Psi}_n(\chi_i)$ can be expressed 
from $\Psi_n(\chi_i)$ with some coordinate changes and 
$\overline{\lambda}_n=\lambda_n$. One can see from Eqn. (\ref{integral}) that 
only the largest eigenvalue ($\lambda_0$) contributes to the partition function
in the thermodynamic limit ($N\to\infty$), i.e. the Gibbs free energy can 
be written as
\begin{eqnarray}
\beta G/N=-\frac{1}{m}\log\lambda_0.
\end{eqnarray}
The equation of state of the system can be obtained from the
relationship between the longitudinal dimension of the channel ($L$)
and the longitudinal pressure, which is given by
$\displaystyle{p = -\frac{1}{H} \frac{\partial F}{\partial L}}$, which corresponds to 
the following equation in the $(N,p,T)$ ensemble:
\begin{equation}
     \rho^{-1}
  =  \frac{\partial G/N}{\partial p},
\end{equation}
where $\rho$ is the density ($\rho^{-1} = HL/N$).
We can also gain some information about the positional distribution of the 
particles from the eigenfunction of the largest eigenvalue, because 
$f(\chi)=\Psi_0(\chi)\overline{\Psi}_0(\chi)$ represents the normalized 
probability distribution function of the cluster with $m$ particles 
($\int d\chi f(\chi)=1$).

The eigenvalue equation (\ref{integral}) simplifies substantially for PHS (see 
equations in Ref. \cite{Gurin3}). For channel-widths ($H$) between $\sigma$ and $2\sigma$ 
Eqn. (\ref{integral}) reduces to the Tonks equation for one-dimensional hard rods 
\cite{Tonks}. In the case of wider channels ($H>2\sigma$), the number of independent positional 
variables of the eigenfunctions goes with $2m-1$ for a cluster of $m$ particles, 
because $m$ coordinates give the positions of $m$ particles along the transverse 
direction, while $m-1$ relative distances are needed for giving the relative 
positions of the particles along the pore. This means that dimers form the clusters 
and $\Psi_0$ depends on two transverse coordinates $(z_1,z_2)$ and one longitudinal 
distance ($x$) for $2\sigma<H<3\sigma$. In general the size of the minimal cluster is equal to 
the number of interacting neighbours of a given particle, which means that $m=1$ 
for first neighbour interactions, $m=2$ for first and second neighbour interactions,
and so on. In general one can conclude that the minimal size of the cluster is equal 
to $i$ and $\Psi_0(z_1,\dots,z_i;x_1,\dots,x_{i-1})$ for $i\sigma<H<(i+1)\sigma$, 
where $i$ is an
integer number. At this point it is important to note that the present transfer 
matrix method cannot be extended for bulk systems, because the number of neighbours of 
a given particle and the size of the minimal cluster diverge with increasing 
channel-width. 
 
\section{Phase behavior and correlations in bulk}

In the present section we describe the bulk phase behaviour of the PHS model using FMT and a Gaussian 
parametrization for the density profile,
\begin{eqnarray}
\rho({\bm r})=\nu \prod_{n=1}^D \left[\sqrt{\frac{\alpha}{\pi}}\sum_{k=-\infty}^{\infty}e^{-\alpha(x_n-kd)^2}\right]
\end{eqnarray}
where $(x_1,x_2)\equiv(x,z)$, $D=1,2$ is the dimensionality of the inhomogeneities ($D=1$ for C and 2 for K phases),
$d$ is the lattice period, i.e distance between layers (for C) or simple square lattice parameter (for K). 
The prefactor $\nu$ represents the occupancy 
probability (one minus the 
fraction of vacancies) for the K phase, while it is equal to $\rho_0 d$ for the C phase, where $\rho_0$ is the mean density per unit cell. 
The free-energy density $\beta {\cal F}[\rho]/A$ is minimised with respect to the Gaussian parameter $\alpha$ and the period $d$ for 
a fixed mean density $\rho_0$. Note that, in the crystal, 
the mean packing fraction and the occupancy probability are related through $\eta_0=\rho_0\sigma^2=\nu\sigma^2/d^2$. 
We should note that the present parametrization 
was recently tested in Ref. \cite{Roij1} where it was compared with the results obtained from the free-minimization of the present functional. 
The authors showed that, apart form minor deviations (small underestimation of the fraction of vacancies and small deviations for the 
predicted C-K coexistence densities), the present approximation works remarkable well. Later in this section the results obtained from the minimization will be compared 
with simulation results of Ref. \cite{Hoover_MD}.

The second purpose of the present section is the study of bulk correlations through the pair correlation function $g({\bm r})$ 
(with ${\bm r}=(x,z)$). We use the FMT functional and follow two different routes.
First, we use the Ornstein-Zernike (OZ) relation, with the direct correlation function obtained from the 
second functional derivative of the FMT functional as an input, and predict the function $g_{\rm fmt-oz}({\bm r})$. Second, we use the test-particle 
route (TP): a hard square is fixed at position $(0,0)$ which acts as an external potential on the rest of the particles. The FMT functional is then minimised 
to obtain $g_{\rm fmt-tp}({\bm r})$. As a third and last step, we have used the Percus-Yevick (PY) approximation which, along with the OZ relation, 
gives an integral equation for the cavity function  $y({\bm r})=g({\bm r})e^{\beta v({\bm r})}$ (with $v({\bm r})$ the hard-core pair potential): 
\begin{eqnarray}
&&y({\bm r})=1+\rho \iint_{A_{\bf 0}} y({\bm r}') d{\bm r}'\nonumber\\
&&-\rho \iint_{A_{\bm r}\cap\left( A\setminus A_{\bf 0}\right)}y({\bm r}')y({\bm r}-{\bm r}')d{\bm r}',
\label{resolver}
\end{eqnarray}
where $\rho$ is the number density, $A$ is the total area of the system, $A_{\bf 0}$ is the area of a square 
of dimensions $\sigma\times\sigma$ positioned 
at the origin, and $A_{\bm r}$ is the area of a square of the same size located at ${\bm r}$. 
Then we can compare the FMT results 
with those from the PY approximation, $g_{\rm py}({\bm r})$. As we will see, the FMT for PHS (and also for hard cubes) 
gives pair-correlations different from the PY result.
This scenario is different for HS, where the direct correlation function obtained from FMT is the same as that from the PY approximation.

Note that the cavity function satisfies the relation
\begin{eqnarray}
y({\bm r})=\left\{
\begin{matrix}
-c({\bm r}), & {\bm r}\in A_{\bf 0}\\
g({\bm r}), & {\bm r}\in A\setminus A_{\bf 0},
\end{matrix}
\right. 
\end{eqnarray}
where $c({\bm r})$ is the direct correlation function. We solved Eq. (\ref{resolver}) numerically for different values of 
packing fractions $\eta=\rho\sigma^2$,
using the fixed-point algorithm. With the function $y({\bm r})$ the inverse structure factor can be obtained by calculating
the integral inside the core of $y({\bm r})$, weighted with cosines functions:
\begin{eqnarray}
&&S^{-1}(q_x,q_z)=1-4\eta \int_0^1 dx \int_0^1 dz \cos(q_xx)\cos(q_z z)\nonumber\\
&&\times y(x,z).
\label{factor}
\end{eqnarray}
The inverse structure factor will be used to look for instabilities of the fluid phase 
with respect to density modulations with crystalline symmetry.  

From $g_{\rm py}({\bm r})$ one can calculate the EOS using the virial route:
\begin{eqnarray}
\beta p_{\rm v}=\rho+\frac{\rho^2}{4}\int_{\cal P} \Delta g(x,z) dl,
\label{virial}
\end{eqnarray}
where the line integral is taken over the perimeter ${\cal P}$ of a square centered at $(0,0)$, while the integrand is the jump of
the pair correlation function at the perimeter (which, according to the PY approximation, is zero inside the core).
Also, the EOS can be calculated from the compressibility route:
\begin{eqnarray}
\frac{\partial \beta p_{\rm c}}{\partial \rho}=1-\rho\hat{c}(0,0)
\label{result}
\end{eqnarray}
where $\hat{c}(0,0)=\int_{A_{\bf 0}} c(x,y) dx dy$ is the Fourier transform of the direct correlation function at zero wave-vector.

Note that $\Delta g({\bm r})=-c({\bm r})$ when ${\bm r}\in{\cal P}$, which follows from the continuity of the cavity function
$y({\bm r})=g({\bm r})-c({\bm r})$ at the perimeter. Substituting this equation in Eqn. (\ref{virial}), together with the FMT expression for the direct 
correlation function,
\begin{eqnarray}
&&-\eta c_{\rm fmt}(x,z)=\left[\xi +\xi^2(2-x-z)\right.\nonumber\\
&&\left.+\xi^2(1+2\xi)(1-x)(1-z)\right]\Theta(1-x)\Theta(1-z), \nonumber\\
&&x,z\geq 0,\quad \xi=\frac{\eta}{(1-\eta)},
\label{fmt}
\end{eqnarray}
one can obtain the analytical result 
\begin{eqnarray}
\beta p_{\rm v}=\frac{\rho}{(1-\eta)^2},
\end{eqnarray}
which coincides with the pressure obtained from the compressibility route (\ref{result}). This result demonstrates that the FMT 
is consistent with respect to the route used.

\begin{figure}
\epsfig{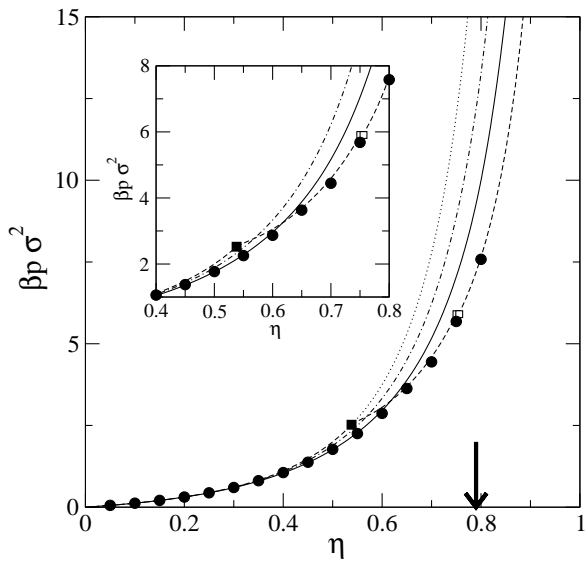}
\caption{EOS of PHS as obtained from the PY virial route (solid curve), compressibility route (dot-dashed curve),
 and FMT approach (dashed curve). The dotted curve shows the EOS of the metastable fluid according to FMT (which coincides with the SPT). 
Filled circles represent MD results from \cite{Hoover_MD},
which predict a fluid-to-crystal transition at a value of packing fraction shown with an arrow. The filled square represents 
the fluid-to-columnar second-order transition, while open squares represent the coexistence densities at the first-order columnar-to-crystal
transition, both predicted by FMT. The inset is a detail of the main figure.}
\label{fig1}
\end{figure}

Fig. \ref{fig1} presents all the EOS obtained from FMT, PY virial route (PYV), PY compressibility route (PYC) and MD simulations 
from \cite{Hoover_MD}. The FMT predicts a bifurcation from the fluid to the C and K phases at the same packing fraction ($\eta=0.538$), with the C phase 
more stable than the K phase up to $\eta\sim 0.75$, at which the K and C free-energy branches cross. The C-K transition is first-order but weak,
with coexisting densities $\eta_{\rm C}=0.726$ and $\eta_{\rm K}=0.730$ obtained from a free minimization (see \cite{Roij1}), while the
corresponding values from a Gaussian parametrization are $0.750$ and $0.756$, respectively. An interesting behaviour of PHS and cubes is 
the relatively large fraction of vacancies (about 15 \% for PHS) that the system can accept at the bifurcation point. Recent MC and MD studies 
show that this is indeed the case; in fact, these 
anisotropic particles (with frozen or free orientations) crystallise with such a large fraction of vacancies (as compared with HS) that it has been 
proposed that the K phase is stabilised by vacancies \cite{Dijkstra1,Marechal1}. By contrast, the FMT does not predict correctly 
the stability of the K phase with respect to the C phase in the range $0.538<\eta<0.750$  
(the C phase should be unstable due to long-wavelength density fluctuations not taken into account in density-functional theory). Also,  
the precise location of the bifurcation point to the K phase is not correct: simulations predict a value about $0.79$, while FMT gives a value 
of $0.538$. However, if we identify the 
C phase as a fluid with lower spatial symmetry, its EOS compares fairly well with that given by MD simulation for  
$\eta>0.6$. Also, the transition to the K phase given by the theory ($\simeq 0.75$) and simulations ($\simeq 0.79$) are similar. 
The major differences are restricted to the interval $0.5 <\eta<0.6$ about the bifurcation point. 
For $\eta>0.75$, above the C-K transition predicted by FMT, the agreement is perfect. This is due to the fact that the FMT recovers 
cell theory, which is known to describe the high-density crystalline phase remarkably well. The PYV EOS 
is very close to that obtained from MD simulation up to $\eta\approx 0.65$, beyond which the pressure is overestimated. 
Also, it exhibits better performance than 
FMT in the interval  $0.5<\eta<0.6$ about bifurcation (see Fig. \ref{fig1}). Finally the EOS from PYC is always above that obtained 
from PYV. Then we can conclude that the PYC EOS is worse than that from PYV compared with MD results. 

\begin{figure}
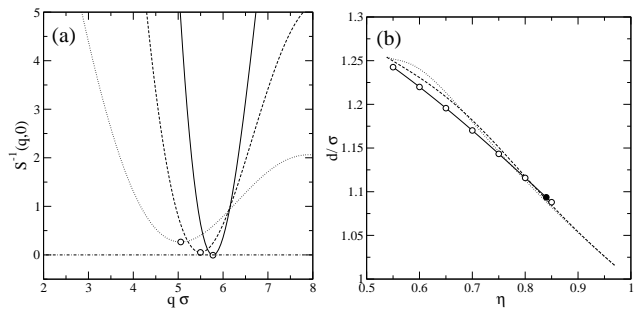

\epsfig{file=fig4a.eps,width=1.6in}
\epsfig{file=fig4b.eps,width=1.6in}
\caption{(a) PY inverse structure factor evaluated in the neighbourhood its absolute minimum
as a function of the wavenumber $q\sigma$ for $\eta=0.55$ (dotted curve), 0.75 (dashed curve) and 0.85 (solid curve).
(b) Lattice parameters $d_m/\sigma$ (with open circles joined by solid lines) 
corresponding to the absolute minima of $S^{-1}(q,0)$ as a function of $\eta$. Filled circle corresponds 
to the estimated fluid-to-crystal transition using the PY approximation. 
Dashed and dotted curves correspond to the periods of columnar and crystal phases, respectively, 
as obtained from the FMT approach.}
\label{fig2}
\end{figure}

Also, we have studied the fluid phase instability with respect to K-phase fluctuations using PYC by calculating the divergence of the inverse 
structure factor calculated from Eqn. (\ref{factor}). Fig. \ref{fig2}(a) shows a zoom of $S^{-1}(q,0)$ close to the value $q^*$ where it becomes zero  
for three different packing fractions. The PY approach predicts an instability at $\eta\approx 0.84$, above that predicted by MD, with 
a value of lattice parameter at bifurcation of $d/\sigma\sim 1.09$. However this result should be taken with some care because at 
these high values of $\eta$ the function $g({\bm r})$ becomes negative at some points (around some of the minima of $g({\bm r})$). 
Fig. \ref{fig2}(b) shows the lattice parameters obtained from FMT 
(for the C and K phases) and the value of $d_m\equiv 2\pi/q_m$, where $q_m$ is the wavenumber corresponding to the absolute minima 
of $S^{-1}(q,0)$ with respect to $q$. We see that the results from both theories are similar in the neighbourhood of the fluid-to-crystal transition predicted by PY.  

\begin{figure}
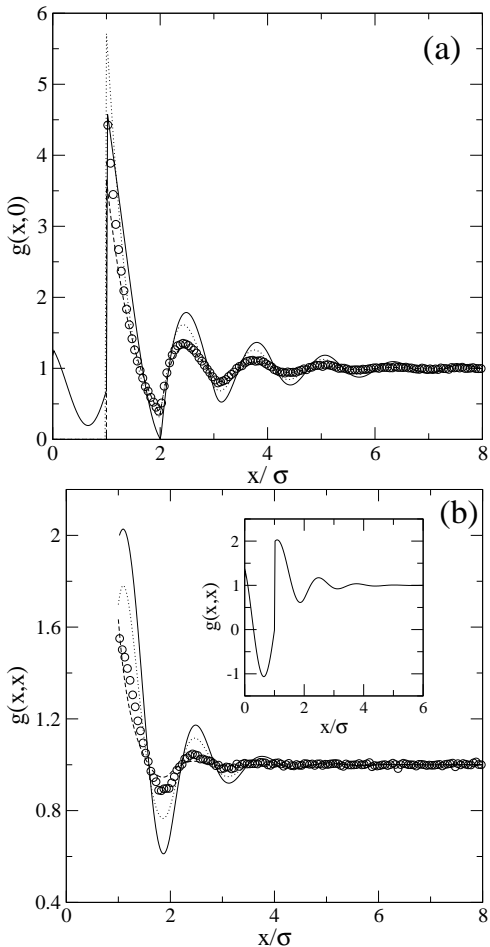

\epsfig{file=fig5a.eps,width=2.5in}
\epsfig{file=fig5b.eps,width=2.5in}
\caption{(a) Pair correlation functions $g_{\rm fmt-oz}(x,0)$ (solid curve),
$g_{\rm py}(x,0)$ (dashed curve), and $g_{\rm fmt-tp}(x,0)$ (dotted curve) for $\eta=0.5$ along the normal direction. 
Symbols show MC simulation results. (b) Pair correlation functions $g_{\alpha}(x,x)$ along the diagonal direction
for the same value of $\eta$, as obtained from the same theories and with meaning of lines and symbols as in panel (a). 
Inset: Complete function $g_{\rm fmt-oz}(x,x)$ including also the core region. }
\label{fig3}
\end{figure}

The pair correlation function $g_{\rm fmt-oz}({\bm r})$ was calculated from FMT using the Eqn. (\ref{fmt}) 
and the OZ  relation. As an example, Fig. \ref{fig3} shows the case $\eta=0.5$ (solid curve) just before the fluid-to-columnar bifurcation point. 
Also, the test-particle route and the PY approximation from Eqn. (\ref{resolver}) were used to calculate $g_{\rm fmt-tp}({\bf r})$ (dotted curve) and 
$g_{\rm py}({\bm r})$ (dashed curve), respectively. MC simulations results are also shown with symbols. 
Panels (a) and (b) show the functions 
$g(x,0)$ and $g(x,x)$, respectively along the normal direction from the centre of the square and along the diagonal direction. 
As can be seen from the figure, the FMT-OZ approximation overestimates bulk correlations, since
the damped oscillations have a larger amplitude and decay more slowly. Similarly, the FMT test-particle route overestimates correlations, but to a
lesser extent. It is interesting to note that although the FMT-OZ approach gives a contact value 
$g_{\rm fmt-oz}(\sigma,0)$ similar to that of simulations, the corner value $g_{\rm fmt-oz}(\sigma,\sigma)$ is overestimated. By contrast,
the test-particle route gives a value $g_{\rm fmt-tp}(\sigma,0)$ 
which overestimates that of simulations, while $g_{\rm fmt-tp}(\sigma,\sigma)$ is very similar to simulations.  
The function $g_{\rm fmt-oz}({\bm r})$ is different from zero inside the core, which shows that the FMT direct correlation function 
is different from that obtained from the PY approximation, a result confirmed from $g_{\rm py}({\bf r})$ (see dashed curve). 
We can see that the PY approximation gives remarkably good results for 
the pair correlation function, except for the value at $(\sigma,0)$, which is underestimated, while $g_{\rm py}(\sigma,\sigma)$ is 
similar to that of simulations. Note that the amplitude and decay of the other peaks 
are very well described by this approximation, at least for this value of packing fraction. We have confirmed 
that $c_{\rm py}({\bf r})$ and $c_{\rm fmt-oz}({\bm r})$ are different, the former being a non-polynomic
function of two variables except for some particular directions for which it becomes a linear or a parabolic function
of a single variable (note that $c_{\rm fmt}({\bm r})$ is a second order polynomial with respect to the variables $x$ and $z$). 
This behaviour is remarkably different from that of hard spheres, 
where the FMT direct correlation function coincides with the PY result. 

\begin{figure}
\epsfig{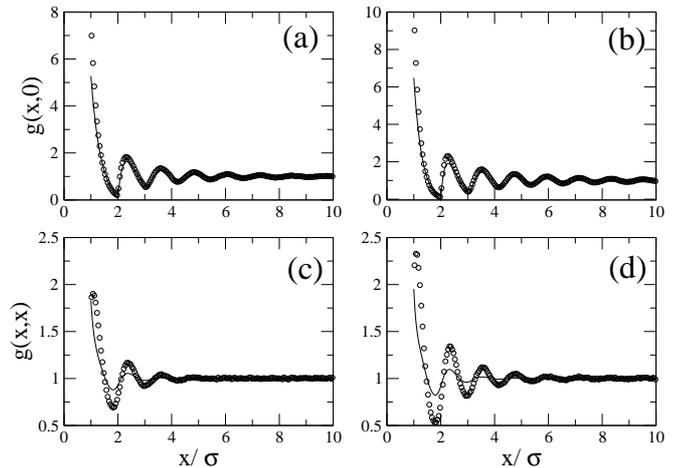}
\caption{Pair correlation functions $g_{\rm py}(x,0)$ [(a) and (b)], and $g_{\rm py}(x,x)$ [(c) and (d)]  for 
$\eta=0.6$ [(a) and (c)], and 0.65 [(b) and (d)]. Symbols are the corresponding MC results. 
}
\label{fig4}
\end{figure}

To finish this section, Fig. \ref{fig4} compares results from the PY approximation and MC simulations 
for $g(x,0)$ and $g(x,x)$ and for packing fractions 
$\eta=0.6$ and $\eta=0.65$ (above the C-K bifurcation predicted from FMT; note that 
the results from FMT are absent in this figure because the Ornstein-Zernike closure 
cannot be used to calculate pair correlations for non-uniform phases).  
At high densities the function $g_{\rm py}(x,0)$ is reasonably close to that from MC (except for the 
contact value). However $g_{\rm py}(x,x)$ strongly underestimates correlations (see Fig. \ref{fig4}). 
To summarize, we can say that the PY approximation, as applied to the calculation of $g({\bm r})$ for PHS and possibly also for 
hard cubes, 
although performing better than FMT, does not have the same degree of accuracy as for HS.  
The fact that it seems to predict a fluid-to-crystal transition 
at densities close to $\eta=0.79$ (in contrast to FMT, which predicts a bifurcation at $0.54$) 
can be used to include it as a main ingredient to construct a modified version of FMT. 
The bad description (with respect to the hard-sphere case)
of correlations featured by the PY theory is a general trend in many models of anisotropic particles.
A modification of FMT to include PY correlations will certainly spoil the accurate description 
of the present density functional for highly confined PHS systems (an issue studied in the next section).  
The $2D\to 1D$ dimensional cross-over property is not fulfilled any more when the structure of the functional 
is modified to include PY correlations.  

\section{PHS confined in channels}

\begin{figure}
\epsfig{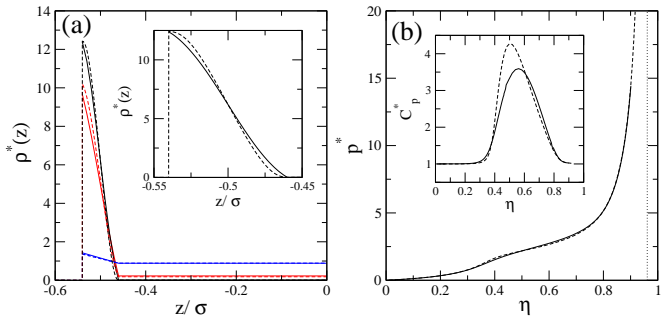}
\caption{(Color online). (a): Normalized density profiles $\rho^*(z)$ of PHS inside the channel with $W=1.08$ as obtained from FMT (dashed) and TMM (solid) 
corresponding to $\eta=0.3$ (blue), 0.6 (red) and 0.8 (black). Inset: Zoom of curves corresponding to $\eta=0.8$. Only the half 
of the density profiles are shown as $\rho(z)=\rho(-z)$ (note that the center of the channel is located at $z=0$). 
(b): EOS of PHS into the channel 
resulting from FMT (dashed) and TMM (solid). The vertical dotted line shows the close-packing value. 
Inset: Heat capacity as a function of $\eta$ from FMT (dashed) and TMM (solid).}
\label{fig5}
\end{figure}

This section is devoted to the effect of high confinement on the thermodynamic and structural properties of fluids of PHS. We will 
evaluate the performance of FMT by comparing the densities profiles, EOS and correlation functions with the exact TMM results and 
with MC simulations
for channels of thickness $H=(W+1)\sigma=2.08$, 2.5 and 2.92. $W$ is defined as the space available to the particle centres of mass
(in units of $\sigma$) in the transverse direction $z$, while the walls of the channel are taken to be parallel to the edge-lengths of the particles
(along the $x$ axis). Later in the section we also present some results for oblique channels 
(with the walls forming $45^{\circ}$ degrees with respect to the $x$ axis). 

\begin{figure}
\epsfig{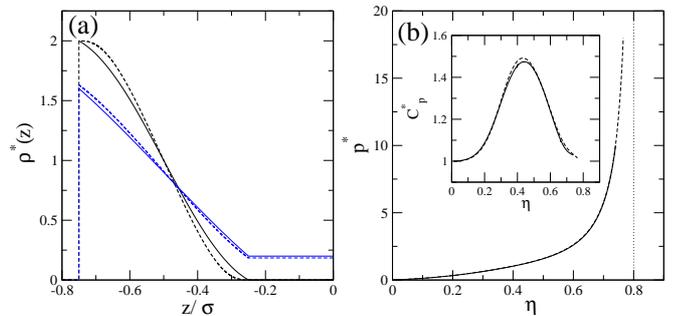}
\caption{(Color online). (a) Normalized density profiles $\rho^*(z)$ of PHS into the channel with $W=1.5$ 
corresponding to $\eta=0.4$ (blue), and 0.7 (black). (b) EOS of PHS into the channel 
resulting from FMT and TMM. Inset: Heat capacity as a function of $\eta$ from FMT and TMM. Lines have the same meaning as in 
Fig. \ref{fig5}}
\label{fig6}
\end{figure}

\begin{figure}
\epsfig{file=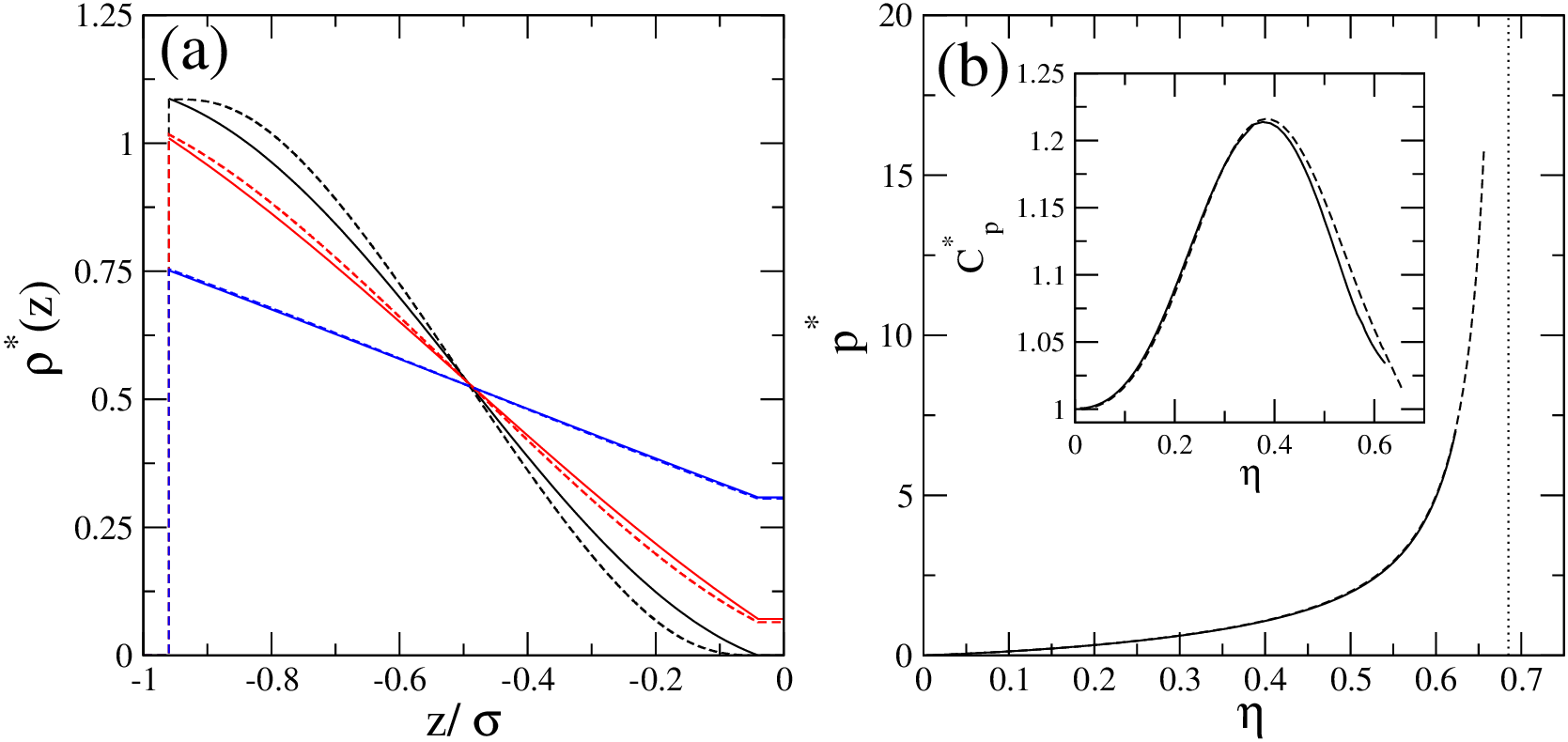,width=3.4in}
\caption{(Color online). (a) Normalized density profiles $\rho^*(z)$ of PHS into the channel with $W=1.92$ 
corresponding to $\eta=0.2$ (blue), 0.4 (red) and 0.6 (black). (b): EOS of PHS into the channel 
resulting from FMT and TMM. Inset: Heat capacity as a function of $\eta$ from FMT and TMM. The lines have the same meanings as in 
Fig. \ref{fig5}}
\label{fig7}
\end{figure}

The grand potential $\Omega[\rho]$ was numerically minimised with respect to $\rho(z)$ by discretising the channel along the transverse 
direction, $ -H/2\le z\le H/2$, using up to $N=H/\sigma\times 10^4$ points, and implementing a conjugate-gradient method.      
Figs. \ref{fig5}, \ref{fig6} and \ref{fig7} show the normalised density profiles $\rho^*(z)=\rho(z)/\int_{-H/2}^{H/2} \rho(z) dz$ [(a) panels], 
the longitudinal pressure in dimensionless units, $p^*\equiv \beta p\sigma^2$ [(b) panels], and the scaled heat capacity at constant 
pressure (without the kinetic term), $C_p^*\equiv C_p/Nk_B-1=(p^*/\eta)^2/(\partial p^*/\partial \eta)$ [inset of (b) panels] 
for different values of packing fraction $\eta$. The latter is calculated as $\eta= \sigma^2H^{-1}\int_{-H/2}^{H/2} \rho(z)dz$ and the 
close-packing value is $\eta_{\rm cp}=2\sigma/H$ (assuming the maximum number of particles that fit in the transverse direction 
is two).

We can see from Fig. \ref{fig5} (a) that the FMT density profiles of confined PHS in a channel 
with $W=1.08$ are very similar to the exact ones. Only small differences are seen in the shape of the
profiles for high packing fractions: while FMT predicts Gaussian-type density profiles (see the inset in (a)), with a small plateau 
near the contact and a Gaussian-like decay, the exact results show density profiles with approximately linear shape, except 
in a very small neighbourhood of the wall. Note that, apart from these differences, intervals in which the density profiles are  
not negligible are exactly the same, leading to a huge adsorption of both layers at the walls.
The EOS predicted from FMT is almost identical to the exact result [see panel (b)]. It is interesting to note that it changes its curvature. This 
behavior is related to the change of structure of confined PHS: while for small packing fractions the system behaves as one-dimensional 
hard segments, as density is increased two highly localized layers of PHS are created near the walls. However, 
no phase transition is apparent in the inset of panel (b): the heat capacity shows no divergence or discontinuity at the packing fraction 
where most structural changes take place inside 
the channel (close to the maximum). We can see that the FMT underestimates the position of the peak 
($\eta_{\rm fmt}\sim 0.5$, while $\eta_{\rm tm}\sim 0.6$) and also 
overestimate its height. 

The results in Figs. \ref{fig6} and \ref{fig7} exhibit the generally
good performance of FMT in describing highly-confined fluid structures and the EOS for wider channels ($W=1.5$ and 1.92). This 
adequacy is due to the dimensional reduction property that the present functional fulfills. Note that now both layers are less 
localized as compared to the $W=1.08$ case.

We have also calculated the pair correlation function $g(0,\tilde{z},x,z)$ of confined PHS. To this purpose we calculated, 
via FMF minimization, the quotient $\rho(x,z)/\rho(z)$, where $\rho(z)$ is the one-dimensional density profile of the 
confined fluid while
$\rho(x,z)$ is the two-dimensional density profile of the confined fluid subjected also to an external potential 
coinciding with that of a fixed PHS of same dimensions positioned at $(0,\tilde{z})$. 

\begin{figure}
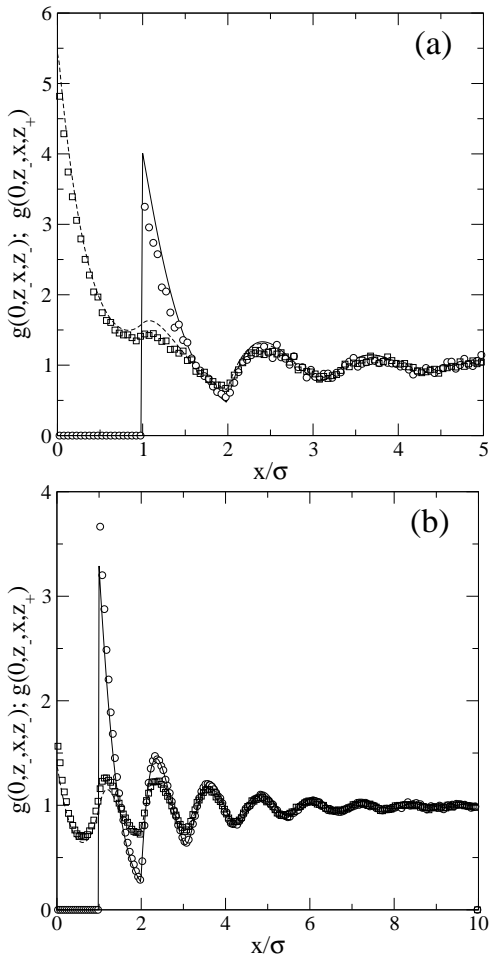

\epsfig{file=fig10a.eps,width=2.5in}
\epsfig{file=fig10b.eps,width=2.5in}
\caption{Pair correlation functions $g(0,z_-,x,z_-)$ (solid line and circles) and $g(0,z_-,x,z_+)$ (dashed line and squares)  for channels with $W=1.08$ and packing fractions 
$\eta=0.4$ (a) and 0.6 (b). Symbols represent MC results. Note that for some range of $x$ the symbols are superimposed over 
the lines showing the good agreement between theory and simulations.} 
\label{fig8}
\end{figure}

\begin{figure}
\epsfig{file=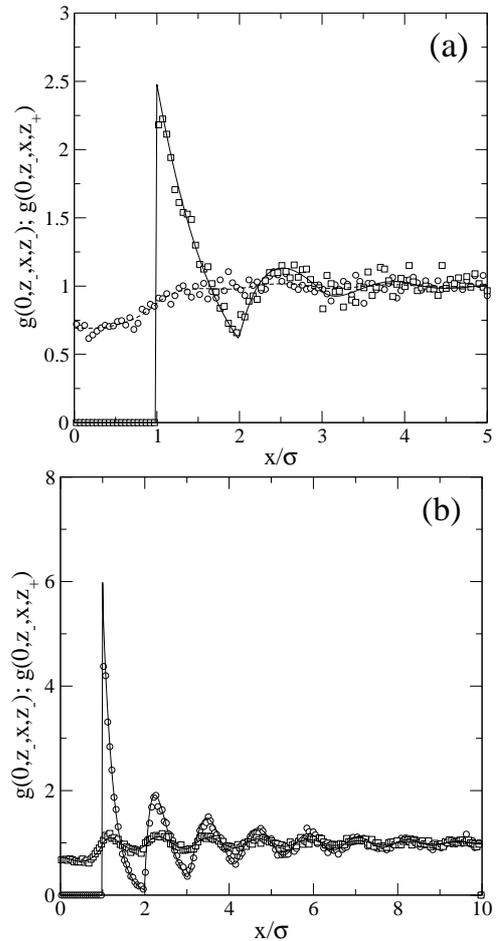,width=2.5in}
\epsfig{file=fig11b.eps,width=2.5in}
\caption{Pair correlation functions $g(0,z_-,x,z_-)$ (solid curve and circles) and $g(0,z_-,x,z_+)$ (dashed curve and squares)  for channels with $W=1.92$ and packing fractions 
(a) $\eta=0.4$ and (b) $0.55$. Symbols represent MC results.}
\label{fig9}
\end{figure}

This function was obtained for channels with $W=1.08$, 1.5 and 1.92 and different values of packing fractions. Also MC simulations 
were carried out to evaluate the performance of FMT on the description of pair correlations. 
The results are shown for the function $g(0,z_-,x,z_{\pm})$ ($z_{\pm}\equiv \pm( H-\sigma)/2$, i.e. one square is fixed at the position $(0,z_-)$ in contact with one of the walls 
while the $z$-coordinate of the other has, or the same $z_-$, or it is located at contact to the other wall at $z_+$.)
in Fig. \ref{fig8} for $W=1.08$ and $\eta=0.4$ [a] and 0.6 [(b)] and in Fig. \ref{fig9} for $W=1.92$ and $\eta=0.4$ [(a)] and 0.55 [(b)].    
Note that the MC data exhibit a large scatter because these data are extracted from a three-dimensional histogram in $(\tilde{z},x,z)$ by 
fixing the bin indexes in two of the dimensions, $(\tilde{z},z)$.

As we can see for thin channels ($W=1.08$) and high enough packing fractions the particles belonging to same or different layers are strongly correlated and the results from FMT 
reproduces those obtained by MC simulations.
Note that the peak at $x=0$ (dashed line) represents a 
strongly correlated dimer where particles are in different  layers.  
When the channel is wide enough ($W=1.92$), correlations between particles belonging to 
different layers decrease dramatically. 
In any case all correlations have an exponential decay, which demonstrates the absence of a phase transition.

The good description of highly confined PHS fluids made by FMT is a consequence of its compliance with 
dimensional crossover. However, we expect that as the channel becomes wider, the high degree of accuracy shown here 
will be lost. To illustrate this, we have looked for possible phase transitions, in wider ($H>3\sigma$) channels,
between confined columnar phases (C$_n$) 
with different number $n$ of layers. A C$_2$-to-C$_3$ layering transition is found at $H/\sigma=3.05$ and bulk
chemical potential $\beta\mu\sim 18$, with coexisting C$_2$ and C$_3$ phases having packing fractions 
$\eta=0.615$ and $0.907$, respectively. Since it is well known that 
confined hard-core interacting particles in dimension $D=1+\epsilon$ (with $\epsilon<1$) 
do not exhibit any phase transitions \cite{Anxo}, this is a spurious prediction of FMT that stems from its mean-field nature.

\begin{figure}
\epsfig{file=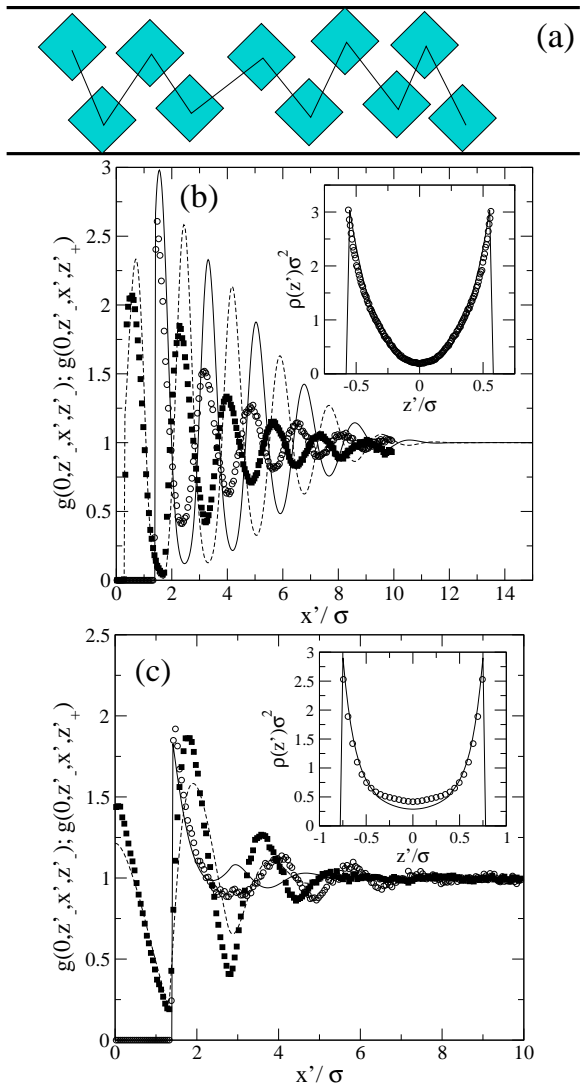,width=3.in}
\epsfig{file=fig12b.eps,width=2.5in}
\epsfig{file=fig12c.eps,width=2.5in}
\caption{(a) Sketch of confined rhombuses in the most likely configuration. (b) Correlation
functions $g(0,z_-,x,z_-)$ (solid) and $g(0,z_-,x,z_+)$ (dashed curve) for the oblique channel with $W=1.13$ and
packing fraction $\eta=0.43$. Inset: Density profile of rhombuses for same $W$ and $\eta$. (c) Same functions as in (b), 
but for $W=1.53$ and
packing fraction $\eta=0.44$. Symbols are MC results [open for $g(0,z_-,x,z_-)$ and filled for $g(0,z_-,x,z_+)$].}
\label{fig10}
\end{figure}

To end this section, we present the FMT results for a system of PHS confined in an oblique 
channel where the walls 
form an angle of 45$^{\circ}$ with respect to the edge-lengths of the squares. 
Note that this system 
is equivalent to that of confined symmetric rhombuses, see sketch in Fig. \ref{fig10}(a). 
To present the results we rotate the reference system 
by 45$^{\circ}$ in such a way that the new coordinates $(x',z')$ obtained after rotation 
are parallel ($x'$) and perpendicular ($z'$) to the walls. Fig. \ref{fig10} (b) shows 
the functions $g(0,z'_-,x',z'_{\pm})$
for the case $W=1.13$ and $\eta=0.43$ for which two rhombuses in contact with opposite 
walls cannot overtake along the channel.
This figure only shows results from FMT and MC simulations (but not from TMM)
because analytic results for the density profile and nonuniform pair distribution function
can only be obtained from TMM when the channel width is such that particles
interact only with nearest neighbours (i.e. with two neighbours). Including
next-nearest neighbours, which is necessary for wider channels, is possible
in principle but a formidable task in practice.
It is clear that the peaks in $g(0,z'_-,x',z'_-)$ and $g(0,z'_-,x',z'_+)$ are out of phase, reflecting the fact that 
particles are positioned in a zigzag configuration (see Fig. \ref{fig10} (a)), which facilitates high-packing configurations in the fluid phase. 
This structure was already observed in \cite{Varga}, where
the TMM was used to study PHS confined in oblique channels and confined hard discs. Despite their high packing 
fraction, rhombuses are distributed over the whole channel, not just next to the walls, as can be seen from the  
relatively high value of the density in the middle of the pore, inset of Fig. \ref{fig10}(b). 
Note that, despite the good agreement between the FMT and MC density profiles, 
correlations are grossly overestimated by FMT.
In Fig. \ref{fig10}(c) the same functions are plotted for the case $W=1.53$ and $\eta=0.44$. 
Now rhombuses in contact with opposite walls can overtake along the channel. 
As pointed out before the TMM cannot be implemented for this case.  
We can see that comparison with MC simulations is reasonable, both for density profile and correlations.
Note that the close-packing density of rhombuses for a given value of $W$ is not as trivially obtained
as for the case of PHS confined in parallel channels (where $\eta_{\rm cp}=2/(1+W)$). 

\begin{figure}
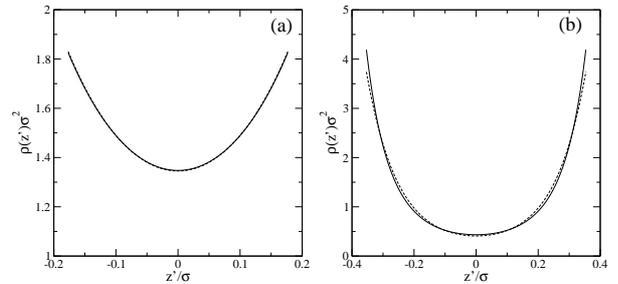

\epsfig{file=fig13a.eps,width=1.55in}
\epsfig{file=fig13b.eps,width=1.5in}
\caption{Density profiles of rhombuses confined in channels with $W=0.354$, and $\eta=0.3$  (a),
 and $W=0.707$, and $\eta=0.4$ (b). 
Solid and dashed curves represent the FMT and TMM results, respectively.}
\label{fig11}
\end{figure}

Finally we compare in Fig. \ref{fig11} the density profiles as obtained from FMT and from TMM for channels with 
$W=0.354$ (a) and $W=0.707$ (b). These channels are sufficiently narrow that the use of the TMM to compute the density profiles is possible.
As we can see the FMT compares perfectly with the exact calculations. In (a) the density profile 
is practicaly constant because we are close to the one-dimensional limit. In (b) the relatively high value of the 
density profile at the centre of the pore reflects the lower particle adsorption at the walls, as compared to the parallel channel.

\section{Conclusions}

In the present article we have compared the results of the FMT functional for the bulk and confined PHS system with
exact TMM results and MD and MC simulations. In the case of the bulk system, the phase behaviour and bulk correlations 
were also tested against a PY approximation. As shown by previous studies \cite{Roij1}, the FMT predicts a spurious 
columnar phase from $\eta>0.538$, which is more stable than the crystal up to densities $\eta\sim 0.75$, 
beyond which the crystal becomes more stable. For larger packing fractions up to the close 
packing limit, FMT agrees extremely well with simulations. Note that simulations 
predict a fluid-to-crystal transition for $\eta\sim 0.79$. Also, if we regard the columnar phase  
as a fluid-like phase with lower symmetries, the FMT equation of state is very close to that from simulations
for densities $0.6<\eta<0.75$. The range 
of packing fractions $0.5<\eta<0.6$ is better described by the virial-route PY equation of state, 
while for $\eta<0.5$ the FMT (which reduces to the SPT for uniform 
densities), the virial- and compressibility-route PY results have a similar 
degree of accuracy. The PY approach predicts a fluid-to-crystal transition at $\eta\sim 0.84$, very similar to that from
simulations.

The bulk pair correlation function $g({\bm r})$ is in general better 
described by the PY approximation, although correlation along diagonal directions are underestimated. 
Bulk pair correlation functions calculated from: (i) the direct 
correlation function from FMT inserted into the OZ equation, and (ii) the test-particle route via functional minimization, both 
overestimate correlations. The first approximation gives a correlation function which is nonzero inside the core, demonstrating that
the direct correlation function given by FMT is different from that obtained from the PY approach. We have shown that, in contrast with 
FMT, the latter does not exhibit in general a polynomial form in the spatial variables.

When PHS are confined in channels so that at most two 
particles can fit along the transverse direction, FMT results, 
as compared with the TMM, have a high degree of accuracy. 
Both theories give quantitatively very similar results for the structure of the fluid (density profiles), the equation of state and 
the constant-pressure heat capacity.  
Also pair correlations obtained from FMT using the test-particle route agree relatively well with simulation results.
For sufficiently narrow channels ($W=1.08$), two inflection points appear in the equation of state which feature the important 
changes in the confined-fluid structure associated with the evolution from a one-dimensional-like fluid to one 
formed by two well-defined layers in contact with the walls. Density 
profiles for high packing fractions are very sharp near the walls. For 
wider channels, density profiles exhibit more fluctuations along the transverse 
direction, and FMT performs extremely well, with heat capacities becoming very similar to the exact results. 

Correlations in the channel are in general well accounted for by FMT. The 
non-uniform pair correlation function for narrow channels at intermediate densities 
reflects a high correlation between particles belonging to a dimer oriented perpendicular to the walls, even higher than for neighbours 
of the same layer. At higher packing fractions  
the situation is the opposite due to a decrease in mean distance between neighbours 
of the same layer, while at the same time neighbours in different layers keep the same distance (see Fig. \ref{fig8}). 
For wider channels neighbours from different layers are always less correlated than those in the same layer (see Fig. \ref{fig9}).
Confinement of rhombuses shows that the most favourable particle configuration is zigzag, and that the fraction of particles 
in the middle of the channel is relatively large despite their high packing fraction. However, in this case,
FMT particle correlations are less accurate than in the case of squares, especially for very narrow
pores.

Finally, we must mention that the FMT functional does not behave correctly as the channel thickness becomes larger and the system approaches the 
two-dimensional limit. This was checked by identifying phase transitions involving two and three layers of squares, a phenomenon not expected for narrow
channels. This feature is a spurious prediction of FMT that reflects the mean-field nature of the density functional.

\acknowledgements
MGP and EV acknowledges the financial support from Grant FIS2013-47350-C5-1-R from Ministerio 
de Educaci\'on y Ciencia of Spain.


\begin{references}
\bibitem{Evans} Evans R. 1990, in \emph{Fundamentals of inhomogeneous fluids} (Edited by
D. Henderson, Mercel Dekker, New York).

\bibitem{Hansen} Hansen J.-P., and McDonald I. R. 2006 in \emph{Theory of simple liquids}
(Academic Press).

\bibitem{Lutsko} Lutsko J. F., and Baus M. 1990 Phys. Rev. Lett. {\bf 64} 761; 1990
Phys. Rev. A. {\bf 41} 6647.


\bibitem{Curtin} Curtin W. A., and Ashcroft 1985 Phys. Rev. A {\bf 32} 2909.

\bibitem{Tarazona1} Tarazona P. 1985 Phys. Rev. A {\bf 31} 2672; 1985 Phys. Rev. A {\bf 32} 3148(E).

\bibitem{Yasha1} Rosenfeld Y. 1989 Phys. Rev. Lett. {\bf 63} 980.

\bibitem{Yasha2} Rosenfeld Y. 1988 J. Chem. Phys. {\bf 89} 4272.

\bibitem{Schmidt1} Rosenfeld Y., Schmidt M., L\"owen H., and Tarazona P. 1997 Phys. Rev. E {\bf 55} 4245.

\bibitem{Tarazona2}Tarazona P. 2000 Phys. Rev. Lett. {\bf 84} 694.

\bibitem{Roth1} Roth R., Evans R., Lang A., and Kahl G. 2002 J. Phys.: Condens. Matter {\bf 14} 12063.

\bibitem{Roth1b} Hansen-Goos H., and Roth R. 2006 J. Phys.: Condens. Matter {\bf 18} 8413.

\bibitem{Roth2} Roth R., 2010 J. Phys. Condens. Matt. {\bf 22} 063102.

\bibitem{Tarazona3} Tarazona P., Cuesta J. A., and Mart\'{\i}nez-Rat\'on Y. 2008.
Density functional theories of hard particle systems.
In \emph{Theory and simulation of hard-sphere fluids and related systems}, ed. A. Mulero, Lect. Notes
Phys. {\bf 753} 247.
{\bf 753} 247.

\bibitem{Cuesta0} Cuesta J. A. 1996 Phys. Rev. Lett. {\bf 76} 3742.

\bibitem{Cuesta1} Cuesta J. A., and Mart\'{\i}nez-Rat\'on Y. 1997 Phys. Rev. Lett. {\bf 78} 3681.

\bibitem{Cuesta2} Cuesta J. A., and Mart\'{\i}nez-Rat\'on Y. 1997 J. Chem. Phys. {\bf 107} 6379.

\bibitem{Yuri1} Mart\'{\i}nez-Rat\'on Y., and Cuesta J. A. 1999 J. Chem. Phys. {\bf 111} 317.

\bibitem{Yuri2} Mart\'{\i}nez-Rat\'on Y. 2004 Phys. Rev. E. {\bf 69} 061712.

\bibitem{Yuri3} Mart\'{\i}nez-Rat\'on Y., Varga S., and Velasco E. 2011 Phys. Chem. Chem. Phys. {\bf 13} 13247.

\bibitem{Velasco1} Velasco E., and Mart\'{\i}nez-Rat\'on Y. 2014 Phys. Chem. Chem. Phys. {\bf 16} 765.

\bibitem{Miguel} Gonz\'alez-Pinto M., Mart\'{\i}nez-Rat\'on Y., and Velasco E. 2013 Phys. Rev. E {\bf 88} 032506.

\bibitem{Harnau} Harnau L., and Dietrich S. 2007 Soft Matter {\bf 3} 156.


\bibitem{Schmidt2} Schmidt M. 2001 Phys. Rev. E {\bf 63} 050201(R).

\bibitem{Schmidt3} Brader J. M., Estermann A., and Schmidt M. 2002 Phys. Rev. E {\bf 66} 031401

\bibitem{Schmidt4} Esztermann A., and Schmidt M. 2004 Phys. Rev. E {\bf 70} 022501;
Esztermann A., Reisch H., and Schmidt M. 2006  Phys. Rev. E {\bf 73} 011409.

\bibitem{Heras} de las Heras D., and Schmidt M. 2013 Soft Matter {\bf 9} 8636.

\bibitem{Yuri_cylinders} Mart\'{\i}nez-Rat\'on Y., Capit\'an J. A., and Cuesta J. A. 2008 
Phys. Rev. E {\bf 77} 051205; Capit\'an J. A., Mart\'{\i}nez-Rat\'on Y., and Cuesta J. A. 
2008 J. Chem. Phys. {\bf 128} 194901.

\bibitem{Mecke1} Hansen-Goos H., and Mecke K. 2009 Phys. Rev. Lett. {\bf 102} 018302. 

\bibitem{Mecke2} Wittmann R., Marechal M. and Mecke K. 2014 J. Chem. Phys. {\bf 141} 064103. 

\bibitem{Mecke3} Wittmann R., and Mecke K. J. Chem. Phys. {\bf 140} 104703. 

\bibitem{Rosenfeld_Mayer} Rosenfeld Y. 1994 Phys. Rev. E {\bf 50} R3318(R).

\bibitem{Marechal1} Marechal M., and L\"owen H. 2013 Phys. Rev. Lett. {\bf 110} 137801.

\bibitem{Mederos} Mederos L., Velasco E., and Mart\'{\i}nez-Rat\'on Y. 2014 J. Phys. Condens. 
Matter {\bf 26} 463101.

\bibitem{Tarazona_0D} Tarazona P., and Rosenfeld Y. 1997 Phys. Rev. E {\bf 55} R4873.

\bibitem{White1} Gonz\'alez A., White J. A., Rom\'an F. L., and Evans R. 1998 J. Chem. Phys. 
{\bf 109} 3637.

\bibitem{White2} Gonz\'alez A.,  White J. A., Rom\'an F. L., and Velasco S. 2006 J. Chem. Phys. 
{\bf 125} 064703.

\bibitem{Wu} Yu Y.-X., and Wu J. 2003 J. Chem. Phys. {\bf 119} 2288.

\bibitem{Mansoori} Kamalvand M., Keshavarzi T., and Mansoori G. A. 2008 International Journal of 
Nanoscience {\bf 7} 245.

\bibitem{Mariani} Mariani N. J., Mocciaro C., Campesi M. A., and Barreto G. F. 2010 J. Chem. Phys. 
{\bf 132} 204104.

\bibitem{Lutsko2} Lutsko J. F. 2006 Phys. Rev. E {\bf 74} 021121.

\bibitem{Hartel} H\"artel A., Oettel M., Rozas R. E., Egelhaaf S. U. , Horbach J., and L\"owen H.
2013 Phys. Rev. Lett. {\bf 108} 226101.

\bibitem{Oettel} Turci F., Schilling T., Yamani M. H., and Oettel M. 2014 Eur. Phys. J. Special 
Topics {\bf 223} 421.

\bibitem{MC} All Monte Carlo simulations were performed on systems with $400-1000$ hard squares, using typically $10^5$ MC steps for
equilibration and $10^6$ MC steps for averaging. The effect of system size was checked by doubling the number of squares at the same
density. Contact values of the correlation functions were extrapolated using fittings to quadratic polynomials with respect to two arguments.

\bibitem{Roij1} Belli S., Dijkstra M., and van Roij R. 2012 J. Chem. Phys. {\bf 137} 124506.

\bibitem{Hoover_MD} Hoover W. G., Hoover C. G., and Bannerman M. N. 2009 J. Stat. Phys. {\bf 136} 715.

\bibitem{Dijkstra1} Smallenburg F., Filion L., Marechal M., and Dijkstra M 2012 Proc. Natl. Acad. Sci. {\bf 109} 17886.

\bibitem{Marechal2} Marechal M., Zimmermann U., and L\"owen H. 2012 J. Chem. Phys. {\bf 136} 144506.

\bibitem{Hoover1} Hoover W. G., and de Rooco A. G. 1962 J. Chem. Phys. {\bf 36} 3141.

\bibitem{Hoover2} Hoover W. G., and Poiriert J. C. 1963 J. Chem. Phys. {\bf 38}, 327.


\bibitem{holandeses1} Rossi L., Sacanna S., Irvine W. T. M., Chaikin P. M., Pine D. J., and Philipse A. P.
2011 Soft Matter {\bf 7} 4139.

\bibitem{holandeses2} Meijer J.-M., Byelov D. V., Rossi L., Snigirev A., Snigireva I. Philipse A. P., and Petukhov A. V
2013 Soft Matter {\bf 9} 10729.

\bibitem{Chaikin} Zhao K., Harrison C., Huse D., Russel W. B., and Chaikin P. M. 2007 Phys. Rev. E {\bf 76} 040401.

\bibitem{Frenkel} Wojciechowski K. W., and Frenkel D. 2004  Comput. Methods Sci. and Tech. {\bf 10} 235.

\bibitem{Torquato} Donev A., Burton J., Stillinger F. H., and Torquato S. 2006 Phys. Rev. B {\bf 73} 054109.

\bibitem{chinos} Zhao K., Bruinsma R., and Mason T. G. 2011 Proc. Natl. Acad. Sci. U.S.A. {\bf 108} 2684.

\bibitem{Escobedo}  Avenda\~no C., and F. A. Escobedo 2012 Soft Matter 8 4675.

\bibitem{Reis} Reis. H., Frisch H., and Lebowitz L. 1959 J. Chem. Phys. {\bf 31} 369.


\bibitem{Tonks}  Tonks L. 1936  Phys. Rev. {\bf 50} 955.

\bibitem{Kofke} Kofke D. A., and Post A. J. 1993 J. Chem. Phys. {\bf 98} 4853.

\bibitem{Joyce} Joyce G. S. 1967 Phys. Rev. {\bf 155} 478; 1967 Phys. Rev. Lett.

\bibitem{Yeomans} Yeomans J. M., \emph{Statistical Mechanics of Phase Transitions}
(Clanderon Press, Oxford, UK, 1992).

\bibitem{Casey} Casey L. M., and Runnels L. K. 1969 J. Chem. Phys. {\bf 51} 5070.

\bibitem{Gurin1} Gurin P., and Varga S. 2011 Phys. Rev. E {\bf 83} 061710.

\bibitem{Varga} Varga S., Ball\'o G., and Gurin P. 2011 J. Stat. Mech.: Theory
Exp. P11006.

\bibitem{Ashwin} Ashwin S. S., Yamchi M. Z., and Bowles R. K. 2013 Phys. Rev. Lett.
{\bf 110} 145701.

\bibitem{Godfrey1} Godfrey M. J., and Moore M. A. 2014 Phys. Rev. E {\bf 89} 03211.

\bibitem{Yamchi} Yamchi M. Z., Aswin S. S., and Bowles R. K. 2015 Phys. Rev. E
{\bf 91} 022301.

\bibitem{Kamenetskiy} Kamenetskiy I. E., Mon K. K., and Percus J. K. 2004
J. Chem. Phys. {\bf 121} 7355.

\bibitem{Gurin2} Gurin P. and Varga S. 2013 J. Chem. Phys. {\bf 139} 244708.

\bibitem{Percus} Percus J. K., and Zhang M. Q. 1990 Mol. Phys. {\bf 69} 347.

\bibitem{Godfrey2} Godfrey M. J., and Moore A. A. 2015 Phys. Rev. E {\bf 91} 022120.

\bibitem{Gurin3} Gurin P., and Varga S. 2015, J. Chem. Phys. {\bf 142} 224503.

\bibitem{Anxo} Cuesta J. A., and S\'anchez A. 2002 J. Phys. A {\bf 35} 2373.

\end{references}
\end{document}